\documentclass[12pt,preprint]{aastex}
\usepackage{graphicx}
\usepackage{lscape}
\usepackage{enumerate}
\usepackage{xcolor}

\usepackage{lineno}

\newcommand{\muvec}{\mbox{\boldmath $\mu$}}

\newcommand{\btheta}{\mbox{\boldmath $\theta$}}
\newcommand{\bdv}[1]{\mbox{\boldmath$#1$}}

\def\btheta{{\bdv\theta}}
\def\bmu{{\bdv\mu}}
\def\bpi{{\bdv\pi}}

\lefthead{JUNG ET AL.}
\righthead{ }

\begin{document}
\title{KMT-2023-BLG-2669: Ninth Free-floating Planet Candidate with $\theta_{\rm E}$ measurements
}

\author{
Youn Kil Jung$^{1}$, 
Kyu-Ha Hwang$^{1}$, 
Hongjing Yang$^{2}$, 
Andrew Gould$^{3,4}$, 
Jennifer C. Yee$^{5}$, 
Cheongho Han$^{6}$, 
Michael D. Albrow$^{7}$, 
Sun-Ju Chung$^{1}$, 
Yoon-Hyun Ryu$^{1}$, 
In-Gu Shin$^{5}$, 
Yossi Shvartzvald$^{8}$, 
Weicheng Zang$^{2,5}$,
Sang-Mok Cha$^{1,9}$, 
Dong-Jin Kim$^{1}$,
Seung-Lee Kim$^{1}$, 
Chung-Uk Lee$^{1}$,
Dong-Joo Lee$^{1}$,
Yongseok Lee$^{1,9}$, 
Byeong-Gon Park$^{1,10}$, 
Richard W. Pogge$^{4}$
}

\affil{$^{1}$Korea Astronomy and Space Science Institute, Daejeon 34055, Republic of Korea}

\affil{$^{2}$ Department of Astronomy, Tsinghua University, Beijing 100084, China}

\affil{$^{3}$Max-Planck-Institute for Astronomy, K\"{o}nigstuhl 17, 69117 Heidelberg, Germany}

\affil{$^{4}$Department of Astronomy, Ohio State University, 140 W. 18th Ave., Columbus, OH 43210, USA}

\affil{$^{5}$ Center for Astrophysics $|$ Harvard \& Smithsonian, 60 Garden St., Cambridge, MA 02138, USA}

\affil{$^{6}$Department of Physics, Chungbuk National University, Cheongju 28644, Republic of Korea}

\affil{$^{7}$University of Canterbury, Department of Physics and Astronomy, Private Bag 4800, Christchurch 8020, New Zealand}

\affil{$^{8}$Department of Particle Physics and Astrophysics, Weizmann Institute of Science, Rehovot 76100, Israel}

\affil{$^{9}$School of Space Research, Kyung Hee University, Yongin, Kyeonggi 17104, Republic of Korea}

\affil{$^{10}$University of Science and Technology (UST), Daejeon 34113, Republic of Korea}

\begin{abstract}
We report a free-floating planet (FFP) candidate identified from the analysis of the microlensing event KMT-2023-BLG-2669.
The lensing light curve is characterized by a short duration $(\lesssim 3\,{\rm days})$ and a small amplitude $(\lesssim 0.7\,{\rm mag})$.
From the analysis, we find the Einstein timescale of $t_{\rm E} \backsimeq 0.33\,{\rm days}$ and the Einstein radius of $\theta_{\rm E} \backsimeq 4.41\,{\mu}{\rm as}$. 
These measurements enable us to infer the lens mass as $M = 8\,M_{\oplus} (\pi_{\rm rel} / 0.1\,{\rm mas})^{-1}$, where $\pi_{\rm rel}$ is the relative lens-source parallax.
The inference implies that the lens is a sub-Neptune- to Saturn-mass object depending on its unknown distance. 
This is the ninth isolated planetary-mass microlens with $\theta_{\rm E} < 10\,{\mu}{\rm as}$, which (as shown by \citealt{gould22}) is a useful threshold for a FFP candidate.
We conduct extensive searches for possible signals of a host star in the light curve, but find no strong evidence for the host. 
We investigate the possibility of using late-time high-resolution imaging to probe for possible hosts. 
In particular, we discuss that for the case of finite-source point-lens FFP candidates, it would be possible to search for very wide separation hosts immediately, although such searches are ``high-risk, high-reward''.
\end{abstract}

{\section{Introduction}
\label{sec:intro}}

In general, free-floating planets (FFPs) refer to planetary-mass objects that are gravitationally unbound to any host. These objects can be formed in protoplanetary disks, like conventional planets, and subsequently ejected by various mechanisms. These ejection processes include dynamical interactions between planets (e.g., \citealt{rasio96, weid96, chatterjee08}), ejections from systems hosting multiple stars (e.g., \citealt{kaib13}), encounters with passing stars (e.g., \citealt{malmberg11}), dynamical interactions within stellar clusters (e.g., \citealt{spurzem09}), or the evolutionary change of the host star after the main-sequence phase (e.g., \citealt{veras2011}).

It is suggested that the majority of planets ejected from their parent systems are terrestrial-mass planets. This is because orbits of inner low-mass planets are more likely to be disrupted by ejection processes, especially by interactions between low- and high-mass (giant) planets. For example, \citet{ma16} found that from population synthesis calculations, typical masses of ejected planets are in the range of $0.3 - 1.0\,M_{\oplus}$. In addition, most of planet ejections occur around massive (FGK-type) stars due to their high probability to host giant planets. \citet{barclay17} reached a similar conclusion using N-body simulations designed for understanding the formation of terrestrial planets around solar-type stars. They found that in such systems, a substantial fraction of protoplanetary materials, including Mars-mass objects, can be ejected in the presence of giant planets.

Free-floating planets can also be formed by gravitational collapse of gas clouds, following a process similar to that of star formation. It is suggested that these processes can extend down to $1-4\,M_{\rm J}$ \citep{whitworth06}. FFPs may also originate from small molecular cloudlets found in H II regions. However, it remains uncertain whether these small clouds can contract \citep{grenman14}.

It would be of theoretical interest to distinguish between FFPs that formed in protoplanetary disks, and those formed by gravitational collapse. This can only be achieved statistically and requires a method that can detect FFPs with a wide range of masses, including those with $M < M_{\rm J}$ that are nearly impossible to be identified with direct imaging. This implies that gravitational microlensing is the only currently available technique for conducting such studies, because it does not rely on detection of photons from a lensing object.

A microlensing event occurs when a foreground object (lens) is aligned with a background star (source). The gravitational field of the lens then distorts and magnifies the light from the source, resulting in a transient increase of the source brightness to an observer. The duration of this brightening, known as the Einstein timescale $t_{\rm E}$, relies on the angular Einstein radius $\theta_{\rm E}$ and the relative lens-source proper motion $\mu_{\rm rel}$: 
\begin{equation}
\label{eq:timescale}
t_{\rm E} = {\theta_{\rm E} \over \mu_{\rm rel}}; \quad \theta_{\rm E} \equiv \sqrt{\kappa M \pi_{\rm rel}},
\end{equation}
where $M$ is the lens mass, $\kappa = 4\,G / (c^2\,{\rm au}) =  8.14\,{\rm mas} / M_{\odot}$, and $\pi_{\rm rel} = 1\,{\rm au}(D_{\rm L}^{-1} - D_{\rm S}^{-1})$ is the relative lens-source parallax. Here, $D_{\rm S}$ and $D_{\rm L}$ are, respectively, the source and the lens distances. As $t_{\rm E} \propto M^{1/2}$, one would expect that microlensing events caused by FFPs are characterized by short timescales ($t_{\rm E} \lesssim 2\,{\rm days}$). However, it is also possible that short timescales are caused by unusually large lens-source proper motions. 

In the cases for which the lens transits, or comes very close to the source, the source size causes deviations in the light curve \citep{gould94, witt94, nemiroff94} relative to its standard point-source/point-lens (PSPL) curve \citep{paczynski86}. These finite-source effects make it possible to measure $\theta_{\rm E}$, which can provide an additional constraint on the lens mass by removing the uncertainty in $\mu_{\rm rel}$. It is expected that many microlensing events by planetary-mass lenses will show strong finite-source effects, because angular radii of sources roughly corresponds to angular Einstein radii of lenses \citep{bennett96}. That is, $\rho = \theta_{*} / \theta_{\rm E} \approx 1$, where $\rho$ is the normalized source radius and $\theta_{*}$ is the angular source radius. Strong finite-source effects extend the duration of an event, in particular for $\rho > 1$, which facilitates the detection of giant source events. This suggests that a promising way to detect FFP candidates is to monitor microlensing events on giant-source stars \citep{kim21,gould22}. The direct lens-mass measurement also requires information on the relative lens-source parallax, but it is difficult to measure $\pi_{\rm rel}$ for short-timescale events \citep{gould13}.

So far, only eight short-timescale finite-source/point-lens (FSPL) events have been identified \citep{mroz18, mroz19a, mroz20a, mroz20b, ryu21, kim21, koshimoto23}. All of their source stars are red giants with the exception of the turn-off star for MOA-9y-5919. In addition, all of these events have $\theta_{\rm E} < 10\,{\mu}{\rm as}$, which is a useful threshold for FFP candidates as shown in Figure 4 of \citet{gould22}. These small $\theta_{\rm E}$ measurements suggest that the lenses are likely to be free-floating and/or wide-orbit planets, because microlensing observations alone cannot definitely rule out the existence of a distant companion star. These planetary-mass objects, combined with statistical studies based on short-timescale events \citep{sumi11, mroz17} and based on FSPL events \citep{gould22, sumi23}, support the presence of a large population of FFPs and/or wide-orbit planets.

Here we present the discovery of a ninth FFP candidate, KMT-2023-BLG-2669, with the measurement of $\theta_{\rm E} \backsimeq 4.41\,{\mu}{\rm as}$. Similar to seven of the eight previous candidates, the source being lensed is a red giant in the Galactic bulge with $\theta_{\rm *} \backsimeq 6.21\,{\mu}{\rm as}$.

{\section{Observations}
\label{sec:obs}}

KMT-2023-BLG-2669 occurred at equatorial coordinates of $({\rm R.A.,}, {\rm decl.})_{\rm J2000} = $(17:44:59.43, $-32$:49:30.65), corresponding to $(l, b) = (-3^\circ\hskip-2pt .39, -1^\circ\hskip-2pt .91)$. The event was found by the Korea Microlensing Telescope Network (KMTNet; \citealt{kim16}). This survey operates three $1.6\,{\rm m}$ telescopes with $2^{\circ} \times 2^{\circ}$ cameras, which are located at the Siding Spring Observatory (KMTA), the South African Astronomical Observatory (KMTS), and the Cerro Tololo Inter-American Observatory (KMTC). The event was in KMTNet field BLG 17, monitored with a cadence of $\Gamma = 1\,{\rm hr}^{-1}$. KMTNet observations were taken in the $I$ band, with a subset of $V$-band observations to measure the source color. These observations were initially reduced using the pySIS package \citep{albrow09}, which employs the difference image analysis (DIA) technique \citep{tomaney96,alard98}.

The event was noticed as a FFP candidate in 2024 March from the investigation of all events identified by the KMTNet EventFinder \citep{eventfinder}, which also included the first discovery of KMT-2023-BLG-2669. For the light curve analysis, the data were then reprocessed using the updated pySIS package developed by \citet{yang24}. We excluded KMTC observations, because they did not cover the magnified part of the light curve and therefore do not play a role in constraining the model.

{\section{Light Curve Analysis}
\label{sec:analysis}}

\begin{figure}
\plotone{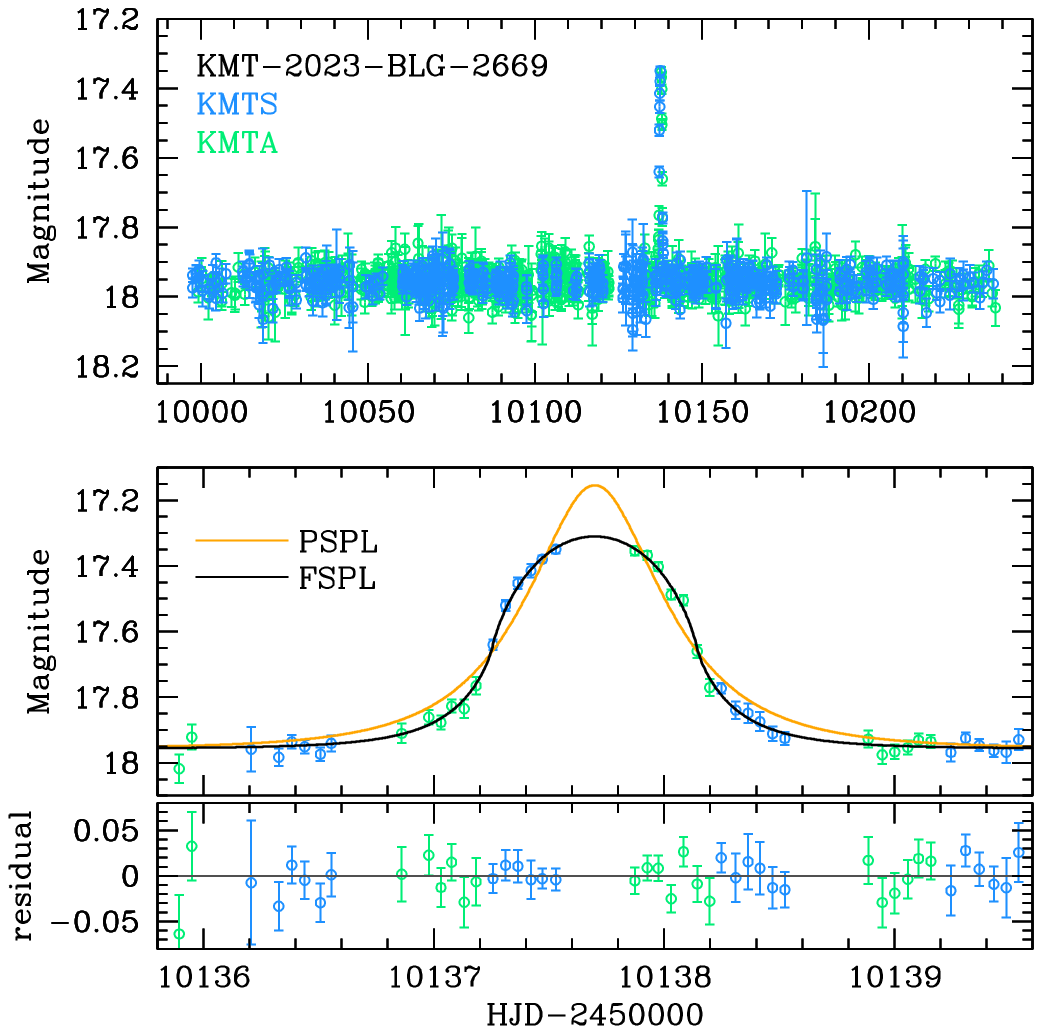}
\caption{Light curve of KMT-2023-BLG-2669. 
The upper panel shows the full data from the 2023 observing season and the lower panel shows the close-up of the event. 
Orange and black curves are the PSPL and FSPL $(f_{\rm B} = 0)$ models, respectively.
}
\label{fig:lc}
\end{figure}

Figure~\ref{fig:lc} shows the data sets obtained from the two observatories (KMTS and KMTA). They show clear deviations from a standard PSPL fit, whose geometry is defined by three parameters $(t_{0}, u_{0}, t_{\rm E})$, where $t_{0}$ is the time of maximal magnification and $u_{0}$ is the impact parameter (normalized to $\theta_{\rm E})$. These deviations can be well explained by the FSPL model, which introduces $\rho$ as a fourth parameter. For each observatory, there are two additional parameters that describe the source $(f_{\rm S})$ and blended $(f_{\rm B})$ flux. To account for the source brightness profile, we adopt a linear limb-darkening profile with $\Gamma_{I} = 0.53$ from the source type (clump giant) estimated in Section~\ref{sec:physical}. In this modeling, the FSPL magnifications are calculated using inverse ray-shooting, and the fit parameters are estimated using the Markov chain Monte Carlo (MCMC) technique.

We fit two models to the light curve, one with a free blended flux $f_{\rm B}$ and the other with fixed $f_{\rm B}=0$. The fit parameters are given in Table~\ref{tab:FSPL}. Our results show that the two models are consistent within $1\,\sigma$, and the model with $f_{\rm B}=0$ is disfavored by only $\Delta\chi^{2} = 0.7$. Hence, the $\chi^2$ difference is perfectly consistent with statistical noise. Moreover, as discussed by \citet{mroz19b}, the blending fractions $\eta=f_{\rm B}/(f_{\rm S}+f_{\rm B})$ for clump giant stars exhibit a bimodal distribution with values close to zero and 1. That is, either the observed flux originates from the source $(\eta \sim 0)$ or the flux is dominated by the blend $(\eta \sim 1)$. Because the two models are all consistent with $\eta \sim 0$, we adopt the fixed $f_{\rm B}=0$ solution.

Our FSPL results suggest that KMT-2023-BLG-2669 appears to be a very low-mass lens, i.e., an FFP candidate. However, if the lens has a host at wide separation, the host may leave a faint signal of its presence in the light curve. To search for such signals, we also fit the light curve using binary-lens/single-source (2L1S) models.

In our 2L1S search, we follow the procedures presented in \citet{kim21}. We introduce three binary-lens parameters to the FSPL solution $(s, q, \alpha)$, i.e., the host-planet separation, the host/planet mass ratio, and the source trajectory angle relative to the host-planet axis. The four parameters $(u_{0}, t_{\rm E}, \rho, s)$ are then scaled to the binary-lens (host+planet) Einstein radius. The coordinate system is centered on the planetary caustic, so that $t_{0}$ is consistent with the value listed in Table~\ref{tab:FSPL}. With these configurations, we define a grid in the binary-lens parameters and run separate MCMC chains with $(s, q)$ held fixed at each grid point. After identifying the local solutions, we then refine the fits using the MCMC, but this time all fit parameters are optimized.

We find no 2L1S solutions with significant $\chi^{2}$ improvement. As shown in Table~\ref{tab:2L1S}, we find two degenerate solutions (Local A and B) differing by $\Delta\chi^{2} = 3.7$. In both solutions, however, the binary-lens parameters are weakly constrained and the fit improvement is insignificant. In particular, the best-fit solution (Local B) is preferred over the FSPL $(f_{\rm B}=0)$ solution by only $\Delta\chi^{2}= 4.7$ with four extra degrees of freedom $(s, q, \alpha, f_{\rm B})$. If we assume perfect Gaussian statistics, this low-level improvement would have a significance of just $p = (1 +\Delta\chi^2/2)\,{\rm exp}(-\Delta\chi^2/2) = 32\%$. In addition, the blended flux for the best-fit solution is negative with the value corresponding to a $\sim 17\,{\rm mag}$ star. Although this solution is mathematically possible (from the fitting process), the negative blending is too large to be caused by normal background fluctuations. Moreover, the best-fit solution is preferred over the solution with $f_{\rm B}=0$ only by $\Delta\chi^2 = 1.2$. Similar to our FSPL results, this low-level $\chi^2$ difference with the large flux errors implies that the negative blending is caused by statistical noise or low-level systematics in the data sets (e.g., \citealt{smith07}). We note that we formally allow $f_{\rm B} < 0$ to avoid a systematic bias in the fit parameters. For the other 2L1S solution (Local A), the source envelops both the planetary and the central caustics, and the caustic signatures are washed out by the strong finite-source effect. This implies that the Local A solution is unlikely because it is essentially identical to the FSPL solution, but with an extra object.

{\section{Physical Parameters}
\label{sec:physical}}

{\subsection{Source Star}
\label{sec:sourcecolor}}

With the measured $\rho$, we estimate $\theta_{\rm E} = \theta_{*}/\rho$ and $\mu_{\rm rel} = \theta_{\rm E}/t_{\rm E}$. For this, we first estimate $\theta_{*}$ using the \citet{yoo04} method of finding the offset between the source and the giant clump (GC) in the color-magnitude diagram (CMD). Typically, the source color is determined from observations in two different bands, achieved by fitting both data sets to the same model or through regression. However, this approach is not feasible in the present case due to the lack of color observations during the event.

\begin{figure}
\plotone{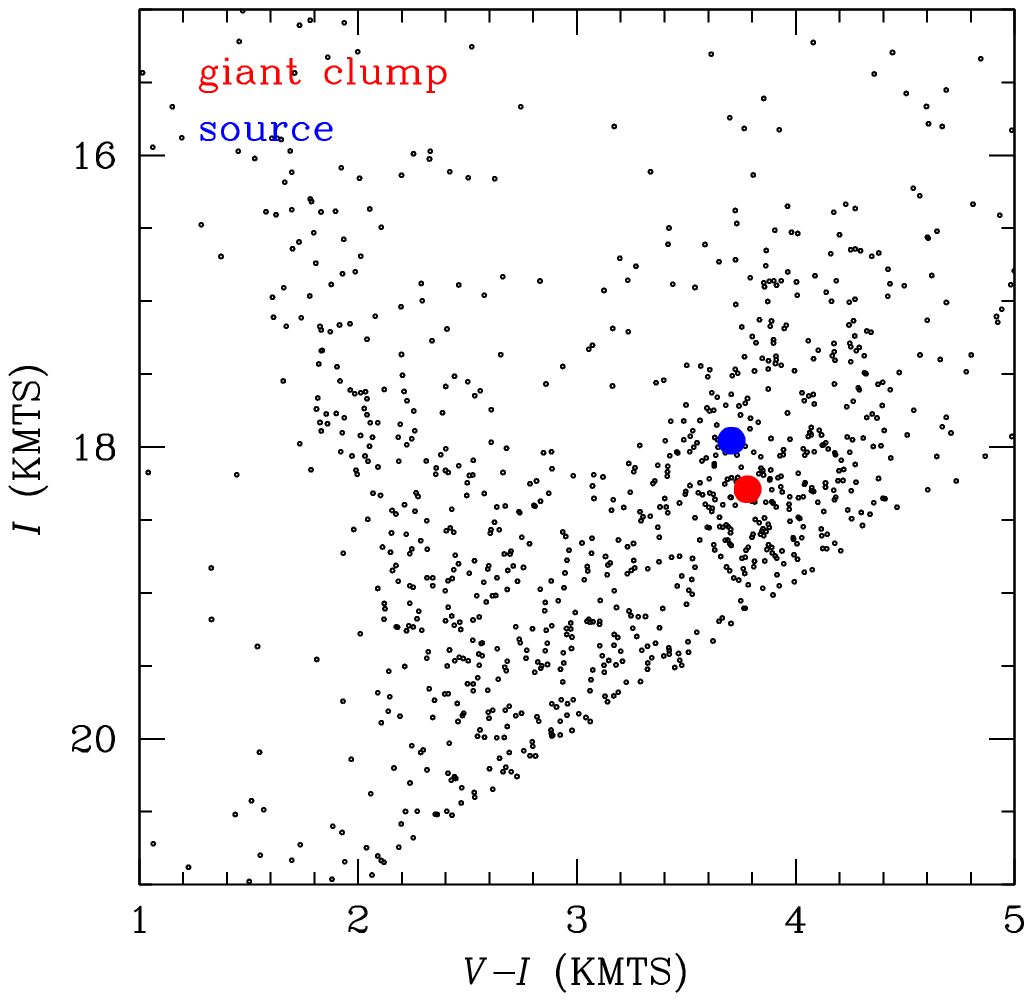}
\caption{KMTS CMD for stars within $4^{\prime}$ of KMT-2023-BLG-2669.
The blue and red dots are the locations of the baseline object (source) and the GC, respectively.
}
\label{fig:KMTScmd}
\end{figure}

Instead, we adopt the color of the baseline object as the source color, given the absence of evidence for blending as shown in Table~\ref{tab:FSPL}. In Figure~\ref{fig:KMTScmd}, we show the positions of the source $[(V-I, I)_{\rm S} = (3.71\pm0.09, 17.96\pm0.01)]$ and the GC $[(V-I, I)_{\rm GC} = (3.78\pm0.03, 18.29\pm0.02)]$ on the KMTS CMD of stars within $4^{\prime}$ of the event. With the measured offset $\Delta(V-I, I) = (-0.07, -0.33)$, we estimate the de-reddened source position by 
\begin{equation}
\label{eq:color}
(V-I, I)_{\rm S, 0} = (V-I, I)_{\rm GC, 0} + \Delta(V-I, I), 
\end{equation} 
where $(V-I, I)_{\rm GC, 0}= (1.06, 14.59)$ is the known de-reddened position of the GC \citep{bensby13,nataf13}. We obtain $(V-I, I)_{\rm S, 0} = (0.99\pm0.09, 14.26\pm0.03)$, which suggests that the source is a K-type red giant.

\begin{figure}
\plotone{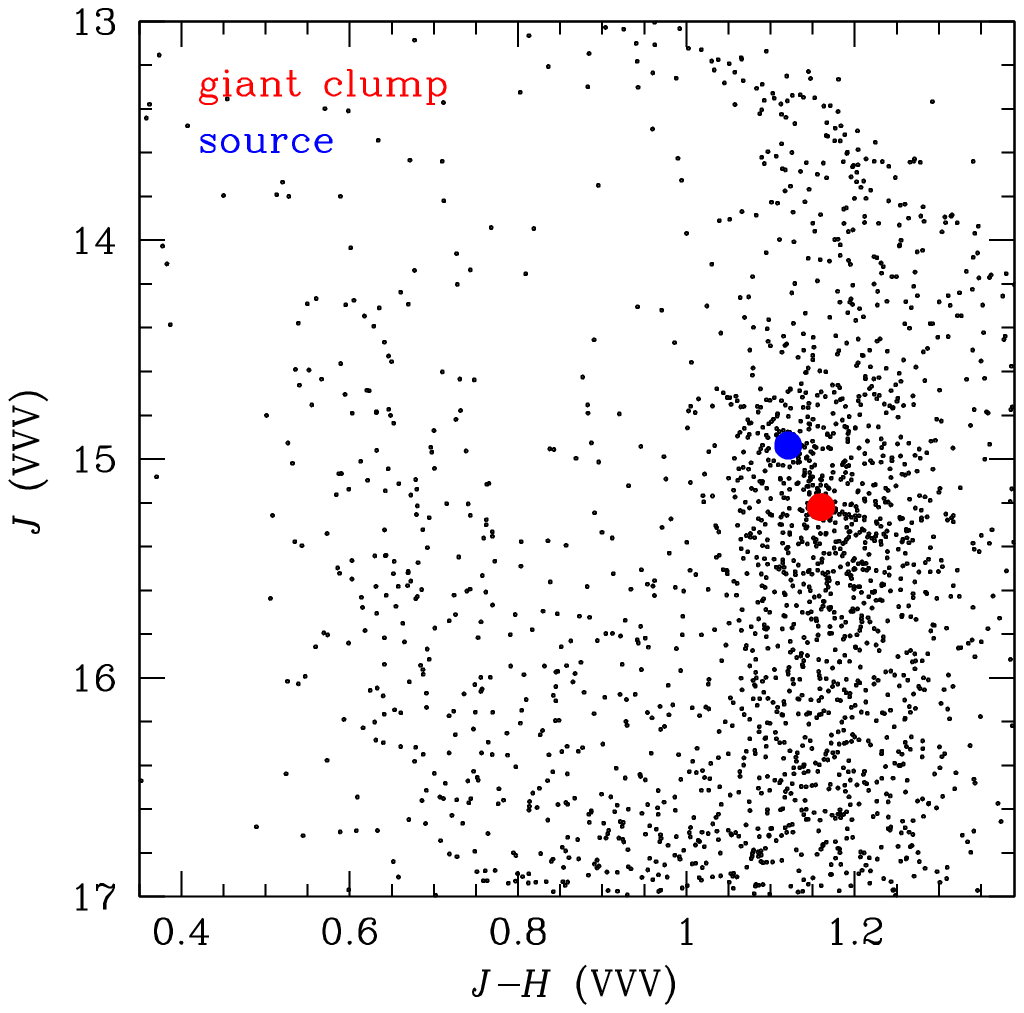}
\caption{VVV CMD for stars within $4^{\prime}$ of KMT-2023-BLG-2669. 
Notations are identical to those in Figure~\ref{fig:KMTScmd}.
}
\label{fig:VVVcmd}
\end{figure}

Despite the bright nature of the source, the uncertainty in $(V-I)_{\rm S, 0}$ is relatively large due to heavy extinction $(A_{I} \sim 3.7)$ toward the event direction. To derive a more precise measurement for the source color, we also find the baseline object (and hence the source) and the GC in the $(J-H, J)$ CMD from the VISTA Variables in the Via Lactea (VVV) survey \citep{minniti23}. As shown in Figure~\ref{fig:VVVcmd}, we find $(J-H, J)_{\rm S} = (1.12\pm0.03, 14.94\pm0.02)$ and $(J-H, J)_{\rm GC} = (1.16\pm0.02, 15.21\pm0.02)$. Using the color-color relations of \citet{bessell88} and Equation~(\ref{eq:color}), we then obtain $(V-I, I)_{\rm S, 0} = (0.99\pm0.04, 14.26\pm0.03)$. The central values are identical to those obtained from the KMTS, but $(V-I)_{\rm S, 0}$ is more precisely constrained. Hence, we adopt this color estimate as the source color.

We then find $[(V-K), K]_{\rm S, 0}$ using again the relations of \citet{bessell88}. Subsequently, we use color/surface-brightness relations of \citet{kervella04}, and finally obtain, 
\begin{equation}
\label{eq:thetastar}
\theta_{*} = 6.21\pm0.38\,{\mu}{\rm as},
\end{equation}
where we add, in quadrature, $5\%$ error to $\theta_{*}$ to address systematic errors (including the adoption of $f_{\rm B} = 0$) in the overall processes. From the $f_{\rm B} = 0$ solution listed in Table~\ref{tab:FSPL}, we then find
\begin{equation}
\label{eq:thetae}
\theta_{\rm E} = 4.41\pm0.29\,{\mu}{\rm as}, \quad
\mu_{\rm rel} = 4.82\pm0.31\,{\rm mas}\,{\rm yr}^{-1}.
\end{equation}
We note that the blended flux can affect the measurements of $\theta_{\rm E}$ only if the source color has been incorrectly estimated \citep{mroz20a,kim21}. That is, if the color had been determined from regression during the event, then even if the blending were different from what we have assumed,
it would not significantly affect the estimate of $\theta_{\rm E}$. However, in our case, the color was estimated from the baseline object, so that if there is (contrary to our arguments) substantial
blending, the color estimate could be incorrect and this would affect $\theta_{\rm E}$.

{\subsection{Proper Motion of the Source}
\label{sec:sourcemotion}}

Because the source is bright and the blended flux is negligible, it is possible to consider the proper motion of the baseline object $\muvec_{\rm base}$ as that of the source $\muvec_{\rm S}$. This enables us to estimate $\muvec_{\rm S}$ from the third Gaia data release (DR3; \citealt{gaia23}). Figure~\ref{fig:gaia} shows DR3 proper motions of stars within $5^{\prime}$ of the event. Red dots correspond to red giant stars (which represent the bulge population), while main-sequence stars (which represent the disk population) are marked in blue. The source proper motion in Galactic coordinates is 
\begin{equation}
\label{eq:gaia}
\muvec_{\rm S}(l, b) = (-8.04, -4.38) \pm (0.39, 0.37)\,{\rm mas}\,{\rm yr}^{-1}.
\end{equation}
In Table~\ref{tab:gaia}, we present the astrometric parameters for the source star from the second Gaia data release (DR2; \citealt{gaia18}) and DR3. We note that the renormalized unit weight error (RUWE) of this measurement is $0.986$, implying that the scatter with respect to the 5-parameter astrometric solution is perfectly consistent with the measurement errors. That is, there is no evidence for astrometric systematic effects. The measurements of $\mu_{\rm rel}$ and $\muvec_{\rm S}$ indicate that the lens proper motion $\muvec_{\rm L}$ is located near the black dashed circle in Figure~\ref{fig:gaia} and is about equally likely to be located in either the disk or the bulge.

{\subsection{Constraints on the Host Star}
\label{sec:sourceconstraint}}

Given that the light curve does not show strong signals for the host, we work to place a lower limit on the projected separation between the planet and a possible host based on the \citet{gaudi00} method. We define a three-dimensional $(101, 61, 37)$ grid in the $(s, q, \alpha)$ plane, which is equally spaced over $0 \leq {\rm log}\,s \leq 2$, $0 \leq {\rm log}\,q \leq 6$, and $0 \leq \alpha \leq 2\pi$. At each grid, we run MCMC chains with fixed $(s, q, \alpha)$ and calculate the $\chi^{2}$ difference $\Delta\chi^{2} = \chi^{2}_{\rm 2L1S} - \chi^{2}_{\rm FSPL}$. For each pair of $(s, q)$, we then estimate the host detection probability from the fraction of light curves that satisfy $\Delta\chi^{2} < \Delta\chi^{2}_{\rm thresh}$. Considering that the 2L1S solution improves the FSPL fit only by $\chi^2_{\rm FSPL} - \chi^2_{\rm 2L1S} = 4.7$, we adopt the threshold of $\Delta\chi^{2}_{\rm thresh} = 25$. From this approach, we find a $90\%$ lower limit on the separation of $s = 1.45$ for possible lens hosts with ${\rm log}\,{q} > 1.7$ (see Figure~\ref{fig:prob}). If the lens host is a $ 0.6\,M_{\odot}$ star, the lower limit of the planet-host projected separation $a_{\bot} = s{\theta_{\rm E,host}}{D_{\rm L}}$ corresponds to $5\,{\rm au}$ and $3\,{\rm au}$ for the disk $(\pi_{\rm rel} = 0.1\,{\rm mas})$ and the bulge $(\pi_{\rm rel} = 0.01\,{\rm mas})$ lenses, respectively.

\section{{Prospects to Search for a Host}
\label{sec:host}}

To date, the hosts of ten 2L1S microlensing planets have been successfully identified in late-time, high-resolution images. 
By comparison, not a single one of the 15 FFPs that have been discovered has yielded a host identification.  
The search for such FFP hosts is important because only by ruling out the existence of a host with high-confidence can we know that the FFP is in fact unbound.
On the other hand, if a host is found, then it will allow us to measure the distance to the FFP, hence an estimate of $\pi_{\rm rel}$, and so an estimate of the FFP mass, $M=\theta_{\rm E}^2/\kappa\pi_{\rm rel}$, as well as the projected separation of this wide-orbit planet.  
Note that even when (contrary to the present case) $\theta_{\rm E}$ is not measured from the event, it can still be determined by measuring the host-source relative proper motion $\mu_{\rm rel}$ from late-time astrometry, thus yielding the mass \citep{refsdal64,gould24}
\begin{equation}
\label{eqn:refsdal64}
M = {\theta_{\rm E}^2\over \kappa\pi_{\rm rel}} = {(\mu_{\rm rel} t_{\rm E})^2\over \kappa\pi_{\rm rel}},
\end{equation}
where $t_{\rm E}$ is the Einstein timescale of the FFP.

At the suggestion of the referee, we present here our analysis of the prospects for identifying (or ruling out) a host of KMT-2023-BLG-2669 in the context of the general problem of identifying hosts for both 2L1S and FFP planets.  
To our knowledge, such an integrated appraisal has not previously been given.

\subsection{{History of 2L1S Host Identifications}
\label{sec:2l1s-host}}

We begin with the published searches for 2L1S hosts. 
Table~\ref{tab:high} summarizes the information derived from a total of 25 published attempted observations of a total of twelve 2L1S lenses. 
It gives the name of the event, the year of observation, the telescope and passband used to observe it, the apparent
magnitudes of the source ($m_{\rm S}$) and lens ($m_{\rm L}$), the contrast ratio $C=10^{0.4|m_{\rm S} - m_{\rm L}|}$, the angular lens-source separation $\Delta\theta$, and the publication reference. 
Figure~\ref{fig:rc} plots the contrast ratio $C$ against the separation $\Delta\theta$, and it also identifies the telescope and passband (Keck $K$, Keck $H$, HST $I$, or HST $V$) by color.

One important ``pattern'' is that all but one of the events occurred more than a decade ago, despite the fact that the total number of published planets has increased by a factor of $>7$ in that decade. 
The reason is that it generally takes a long time for the source and lens to separate sufficiently to resolve the two, given present technology. 
In particular, in a 2L1S event, the host is known to be very close to the source at the time of the event, typically $\Delta\theta\la 1\,{\rm mas}$, so that it is clearly impossible to separately resolve the host and the source at that time. 
This may appear to be an obvious point but it will play an important role.

The second ``pattern'' is that, as one would expect, larger separation permits detections at higher contrast ratio. 
That is, Figure~\ref{fig:rc} shows an upper envelope of increasing contrast ratio with increasing separation. 
There appear to be three exceptions to this pattern, OGLE-2016-BLG-1195 and the HST $V$-band observations of MOA-2008-BLG-379 and OGLE-2012-BLG-0950, each of which has an anomalously high contrast ratio for its separation. 
However, \citet{gould23} showed that for OGLE-2016-BLG-1195, the high-contrast star identified by \citet{vandorou23} is not actually the lens because they misidentified the source star of the underlying microlensing event. 
For the HST $V$-band observation of MOA-2008-BLG-379, the measured lens flux based on the coordinate transformation between HST and Keck images is highly uncertain $(f_{{\rm HST}, V, {\rm L}} = 50 \pm 42)$ and the lens magnitude is only marginally detected at $\sim 1\sigma$ significance \citep{bennett24}.
Regarding the HST $V$-band observation of OGLE-2012-BLG-0950, it is a real detection, which does confirm the HST $I$-band and Keck $K$-band observations that are shown at similar separations in Figure~\ref{fig:rc}. 
However, when this detection is considered by itself (Figure 2b of \citealt{bhattacharya18}), the $68\%$ error contour spans a factor $2$ in lens flux and a factor $1.4$ in separation. 
Hence, it is not clear that it would have been considered a reliable detection without ``prior knowledge'' from other observations.

\subsection{{History of FFP Host Searches}
\label{sec:ffp-host}}

To date, the only published search for FFP hosts is the one carried out by \citet{mroz24}, who did late-time Keck $H$-band imaging of the six FFPs that had been discovered by \citet{mroz17}. 
These were taken $6.0$ to $10.1$ years after the events.

Before continuing, we point out that because all six of these FFPs were PSPL events, none had measurements of $\theta_{\rm E}$ (or $\mu_{\rm rel} = \theta_{\rm E}/t_{\rm E}$). 
This fact will become important when we discuss the problem of verifying FFP hosts in Section~\ref{sec:verify}, which in turn impacts the decision of when to take the first high-resolution observation.

The astronomical techniques to search for FFP hosts are essentially identical to the ones described above for 2L1S hosts, but the physics can be substantially different. 
The key point is that for 2L1S events (as mentioned above) one knows for a fact that $\Delta\theta\la 1\,{\rm mas}$ at the time of the event ($t=t_0$), so there is no point in searching for hosts immediately. 
By contrast, for FFP events, one has no idea whether $\Delta\theta\la 1\,{\rm mas}$ (as above), $\Delta\theta\sim 40\,{\rm mas}$, $\Delta\theta\sim 100\,{\rm mas}$, or $\Delta\theta\rightarrow \infty$ (unbound FFP). 
In the first case, one must wait for the host and source to separate (just as for 2L1S). 
In the fourth case, one will never find the host regardless of how long one waits. 
However, consider Figure~\ref{fig:rc}. In either the second or third case (wide-orbit planet), the host will be separated from the source at $t_0$. 
Therefore, it is potentially detectable at $t_0$ depending on the concrete circumstances of the event.

\citet{mroz24} evaluated the results of their search within the framework of the first option, that is, $\Delta\theta(t_0)\la 1\,{\rm mas}$, so that $\Delta\theta$(now)$=\mu_{\rm rel}\Delta t$ (with no significant offset due to $a_\perp$). 
In practice, they simulated possible host stars with $\Delta\theta(t_0)=0$, but their results would have been nearly identical for any value of $\Delta\theta(t_0) \ll {\rm FWHM} \sim $ 54--84 mas, for example, $\Delta\theta(t_0) = 10\,{\rm mas}$. 
For this example, and under the assumption $D_{\rm L} = 4\,{\rm kpc}$, this would correspond to $a_\perp = 40\,{\rm au}$, similar to Neptune, which is a very plausible source of the FFPs assuming they are bound. 
On the other hand, given our almost complete lack of information, the FFPs might be wide-orbit planets at $a_\perp=300\,{\rm au}$, in which case $\Delta\theta(t_0) = 75\,{\rm mas}$ at $D_{\rm L} = 4\,{\rm kpc}$ or $\Delta\theta(t_0) = 150\,{\rm mas}$ at $D_{\rm L} = 2\,{\rm kpc}$. 
Such hosts would have been at least as detectable in images taken at the time of the event as the simulated hosts shown in their Figure~1 were from the actual late-time observations. 
And such putative hosts would still be detectable in the late-time images because the lens-source vector separation would only have changed by of order $60\, {\rm mas}$. 
Thus, considering that no stars were detected $\Delta\theta < 110\,{\rm mas}$ in any of the five cases of successful observations, one could rule out various large $\Delta\theta(t_0)$ systems as well. 
However, as we discuss in Section~\ref{sec:verify}, if there had been any detections of large $\Delta\theta(t_0)$ systems, it would have been essentially impossible to verify them. 
Indeed, \citet{mroz24} did detect a total of $4$ stars with $\Delta\theta<200\,{\rm mas}$. 
These are most likely random field stars, but if they were true hosts, then this fact could only be verified for the cases with $\Delta\theta(t_0)\la\,{\rm FWHM}$. 
See Section~\ref{sec:verify}.

While Figure~1 of \citet{mroz24} is not expressed in terms of contrast ratios, one can combine it with their Table~1 to derive contrast-ratio limits for the $5$ events at, for example, $\Delta\theta=(50,100,200)\,{\rm mas}$. 
We originally made such estimates. 
However, P.~Mr\'oz (2024, private communication) kindly sent us corrected values based on the original files. 
For the five events (in the order of the figure panels), these are
$C_{\rm max}(50\,{\rm mas}) = (7.8, 4.4, 2.8, 5.1, 5.4)$, 
$C_{\rm max}(100\,{\rm mas}) = (24,12,25,16,12)$, and 
$C_{\rm max}(200\,{\rm mas}) = (44,26,93,25,29)$.
The first of these groups is in good agreement with the upper envelope position at about the same separation for MB09319 in Figure~\ref{fig:rc}. 
The second shows that the upper envelope at MB07400 is an underestimate by a factor $\sim 1.5$, probably due to sparse data at these separations. 
The last shows that contrast ratios corresponding to about $4\,{\rm mag}$ can be achieved at $200\,{\rm mas}$ using Keck.

\subsection{{Verification}
\label{sec:verify}}

In principle, there are three ways that a candidate host could be verified. 
All three require measuring the vector candidate-source offset, $\Delta\btheta(t)$, at two epochs $t_1$ and $t_2$, which then yields the candidate-source proper motion, $\bmu_{\rm c-s} = (\Delta\btheta(t_2)-\Delta\btheta(t_1))/(t_2-t_1)$. (Note that if $\mu_{\rm c-s}$ is inconsistent with zero, this automatically rules out that the candidate is a companion to the source.) 
The first method is to demonstrate proximity, that is, show that the candidate-source separation $\Delta\btheta(t_0) = \Delta\btheta(t_1) - \bmu_{\rm c-s}(t_1 -t_0)$ is small compared to what could plausibly be generated by a random star. 
The second method is to show that the amplitude of candidate-source relative proper motion $\mu_{\rm c-s} = |\bmu_{\rm c-s}|$ is compatible with the value of $\mu_{\rm rel}$ measured during the event. 
And the third method is to show that the direction ${\hat\bmu}_{\rm c-s}\equiv \bmu_{\rm c-s}/\mu_{\rm c-s}$ is compatible with direction of the microlens parallax vector $\bpi_{\rm E}$ as determined during the event.

The third of these will be extremely rare\footnote{If we exclude space-based observations, there are only two known ways to measure microlens parallax: annual parallax \citep{gould92} and terrestrial parallax \citep{holz96,gould97}. FFP events are too short to be probed by annual parallax, while \citet{gould13} showed that terrestrial-parallax measurements will be
extremely rare.} for FFPs until there are space-based FFP surveys. 
The second is possible for FSPL FFPs, which includes KMT-2023-BLG-2669 but does not include any of the \citet{mroz24} events, so we return to this case below.

Thus, the only way to verify candidates for these PSPL events is to demonstrate that $\Delta\theta_{\rm c-s}(t_0)$ is small (compared to expectations from random field stars). 
This immediately implies that it is impossible to search for very wide separation planets in PSPL FFP events (unless there is a $\bpi_{\rm E}$ measurement) because none of the three paths of verification are available for such events. 
Hence, the approach of \citet{mroz24}, which ignored this possibility, was appropriate.

\subsection{{Application to KMT-2023-BLG-2669}
\label{sec:app-to-kmt}}

By far the closest analog to KMT-2023-BLG-2669 among the \citet{mroz24} FFPs is OGLE-2012-BLG-1396: both sources are about $0.3\,{\rm mag}$ above the clump and both suffer about $A_K = 0.6$ magnitudes of extinction \citep{gonzalez12}. 
The one major difference is that KMT-2023-BLG-2669 has measurements for $\theta_{\rm E}$ and $\mu_{\rm rel}$
($\theta_{\rm E} = 4.4\pm 0.3\,{\rm mas}$, $\mu_{\rm rel} = 4.8\pm 0.3\,{\rm mas}\,{\rm yr}^{-1}$), while
OGLE-2012-BLG-1396 does not.  This difference has two important implications, which we touch on below. 
Thus, the OGLE-2012-BLG-1396 panel of Figure~1 from \citet{mroz24} can be used to understand prospective high-resolution imaging of KMT-2023-BLG-2669, albeit with some modification. 
We refer the reader to this panel.

The panel shows a cloud of magenta points that represent an ensemble of simulated hosts, centered at $\Delta\theta \sim 60\,{\rm mas}$ and $H = 22$, which is $\sim 7\,{\rm mag}$ below the $\Delta\chi^2 = 500$ detection threshold. 
As a result, by appearance, the overwhelming majority of the cloud lies below this threshold. 
Somewhat contrary to this appearance, their Table~1 says that $10.7\%$ of these magenta points would have been detected. 
However, P.\ Mr\'oz (2024, private communication) clarified that whereas their simulated bulge lenses were drawn from a \citet{chabrier03} initial mass function (IMF) restricted to $M < 1\,M_\odot$, the disk lenses were drawn from an unrestricted
\citet{chabrier03} IMF. 
If the IMF is restricted to, say, $M < 1.3\,M_\odot$ in order to mimic a disk present-day mass function, then the overall fraction of recovered hosts falls to about $2\%$.

The OGLE-2012-BLG-1396 cloud is dispersed by a factor $\sim 2$ in the separation direction, which is a consequence of $\mu_{\rm rel}$ not being measured. 
For KMT-2023-BLG-2669, the cloud would be very similar in the vertical direction but much more compact in the horizontal direction ($\sigma(\mu_{\rm rel})/\mu_{\rm rel} = 6.5\%$) due to the fact that $\mu_{\rm rel}$ is measured. 
Even waiting $40$ years (2063), at which time the cloud would be at $\Delta\theta = 193\,{\rm mas}$, its center would still lie $3.1\,{\rm mag}$ below the detection threshold. 
This is similar to the offset shown in the OGLE-2015-BLG-1044 panel, for which their detection estimate was $26\%$ of possible host stars.
Applying the same $(9\%)$ correction estimated above for OGLE-2012-BLG-1396, this should be reduced to $17\%$. 
However, long before this $40$ years has elapsed, adaptive optics (AO) will be available on much larger telescopes, and the feasibility of using these telescopes should be assessed rather than waiting for Keck.

\subsection{{High Risk, High Reward}
\label{sec:highrisk}}

Finally, we should assess the possibility of detecting the host of KMT-2023-BLG-2669 for the case of very wide orbit planets ``immediately'' (within a few years after the event), which could be possible in principle due to the fact that $\mu_{\rm rel}$ is measured. 
That is, some fraction of random field stars that might be present at large separations would be rejected because the resulting $\mu_{\rm c-s}$ proper motions would be inconsistent with what we have measured from the event, which has a 2-$\sigma$ range of $4.2\,{\rm mas}\,{\rm yr}^{-1} <\mu_{\rm rel} < 5.4\,{\rm mas}\,{\rm yr}^{-1}$. 
From Figure~\ref{fig:gaia}, roughly $20\%$ of random field stars would be compatible with this constraint. 
This would appear to be a modest enhancement (relative to no $\mu_{\rm rel}$ measurement). 
However, if a star were detected at, for example, $\Delta\theta \sim 200\,{\rm mas}$, it would have to be brighter than $H < 19$, and the expected number of such stars within a $200\,{\rm mas}$ circle is only about $0.05$. 
In this context, the extra factor $\sim 5$ vetting from the $\mu_{\rm rel}$ measurement would be significant.

At present we have absolutely no idea of whether there are such very wide separation planets. 
For example, an early M dwarf at $2\,{\rm kpc}$ would have $H \sim 18.3$ and would thus be detectable. 
Its projected separation would be $a_\perp= 400\,{\rm au}$. 
Its distance could be measured photometrically with additional observations so that the lens-source relative parallax $\pi_{\rm rel} \simeq 0.37\,{\rm mas}$ could be estimated as well. 
Then $M = \theta_{\rm E}^2/\kappa\pi_{\rm rel} \simeq 2\,M_\oplus$. 
While the chance of such a discovery is small, a success would probe a region of parameter space about which we currently have essentially no information. 
Therefore, in our view, such observations should be put in the high-risk high-reward category.

A similar logic is likely to apply to many of the seven other giant source FFPs with measured proper motions.

\section{Summary}

\begin{figure}
\plotone{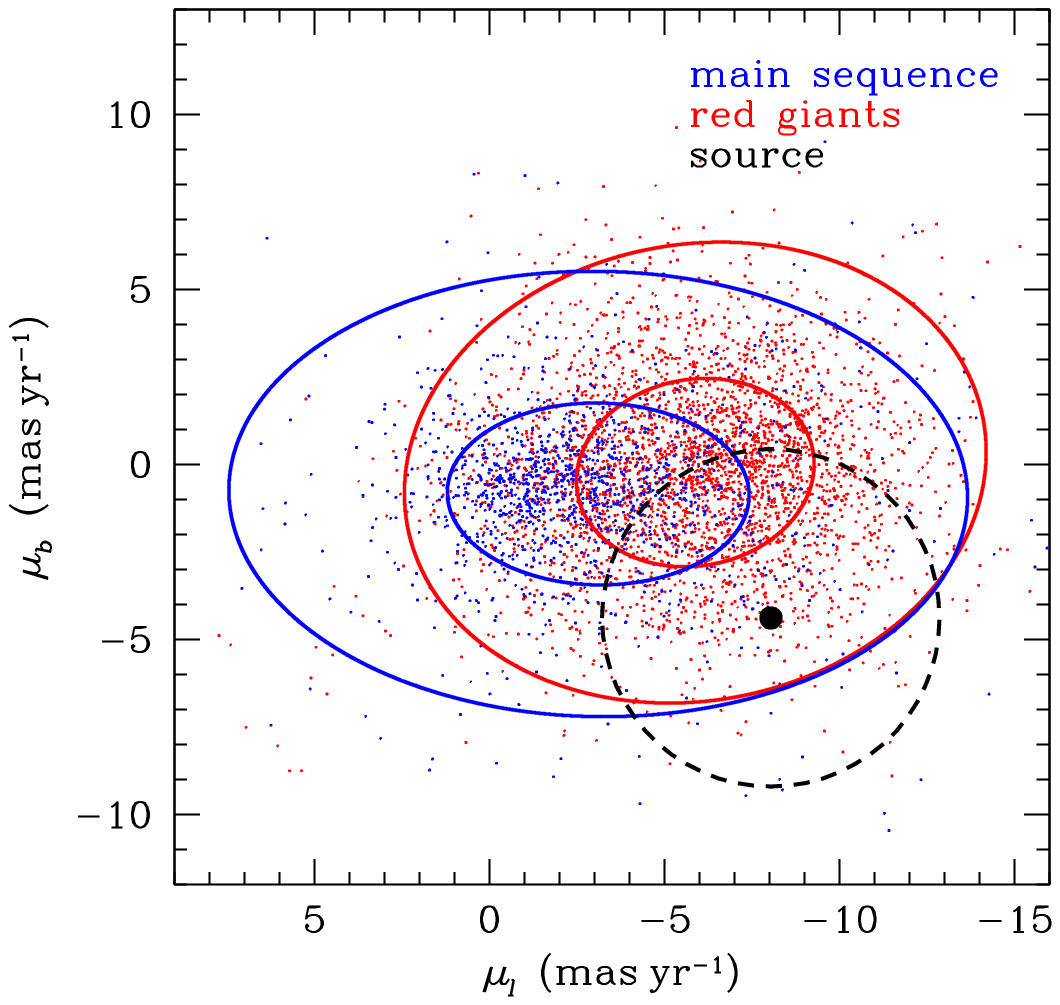}
\caption{Gaia DR3 proper motions of stars within $5^{\prime}$ of KMT-2023-BLG-2669. 
Blue and red dots correspond to red giant and main-sequence stars, respectively. 
In each population, two solid contours include $68\%$ and $95\%$ of all stars. 
The black dot is the location of the baseline object (source). 
The dashed circle is the locus of lens-source relative proper motions, $\mu_{\rm rel} = 4.82\,{\rm mas}\,{\rm yr}^{-1}$. 
}
\label{fig:gaia}
\end{figure}

\begin{figure}
\plotone{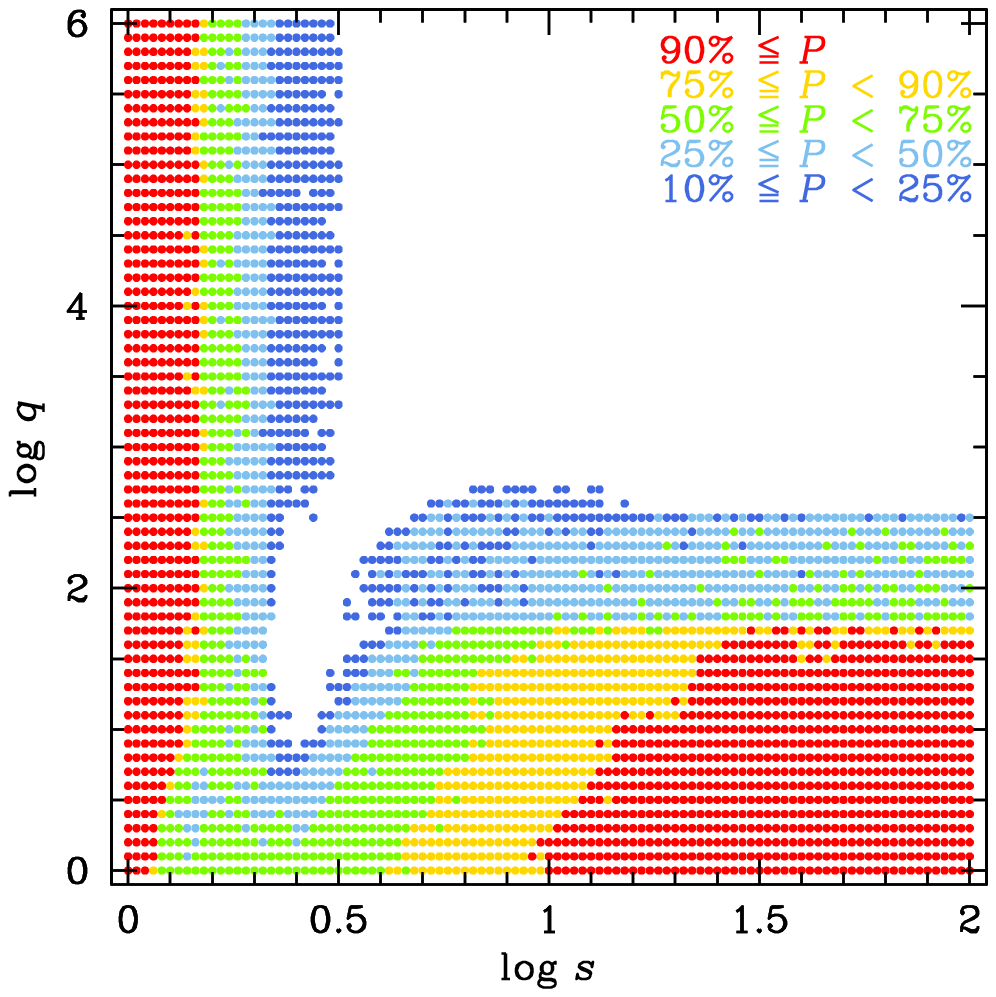}
\caption{Detection probability for the putative host star in the $(s, q)$ plane. 
}
\label{fig:prob}
\end{figure}

\begin{figure}
\plotone{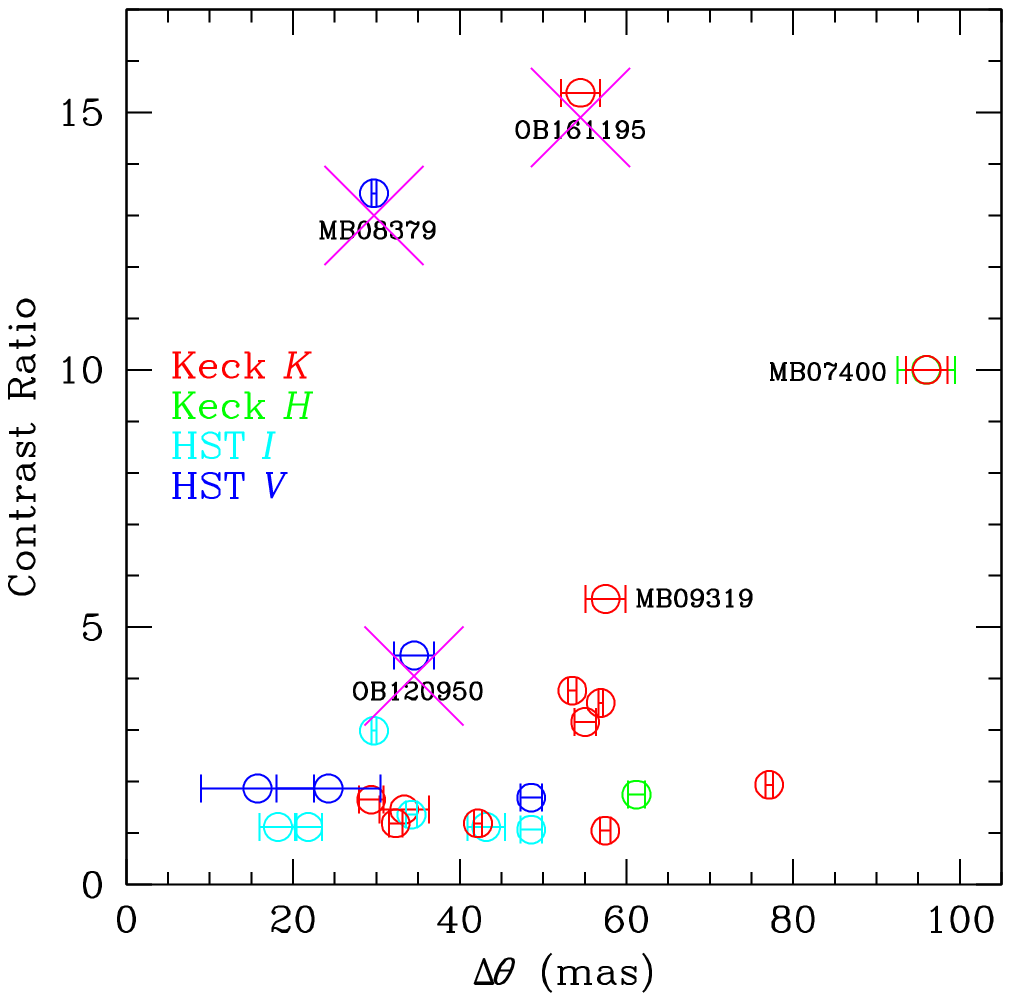}
\caption{Contrast ratio versus host-source separation for a total of 25 late-time observations of a total of twelve 2L1S microlensing events. 
The observatory and filter of each measurement is specified by its color, as indicated in the legend. 
There is generally an envelope of rising maximum contrast ratio as a function of separation, but with three exceptions. 
First, the high contrast ratio for MB08379 (HST $V$-band) would not be considered as a reliable detection, because the measured lens flux is highly uncertain. 
Second, the anomalously high contrast ratio for the OB120950 (HST $V$-band) observation, when considered independently of the HST $I$-band and Keck $K$-band observations, has a factor $2$ range (at 1-$\sigma$) in lens flux, so it is not clear that it would be regarded as a detection by itself. 
Third, the high-contrast star that was apparently detected for OB161195 is known not to be the lens because the observers investigated the wrong source star.
}
\label{fig:rc}
\end{figure}

KMT-2023-BLG-2669 is well described by an FSPL model with a timescale of $t_{\rm E} = 0.334\pm0.014\,{\rm days}$ and an Einstein radius of $\theta_{\rm E} = 4.41\pm0.29\,\mu{\rm as}$. These measurements imply that the lens is an FFP candidate, although its mass is not unambiguously determined due to the lack of $\pi_{\rm rel}$ measurement:
\begin{equation}
\label{eq:lensmass}
M = {\theta_{\rm E}^{2} \over \kappa\pi_{\rm rel}} = 8\,M_{\oplus} {0.1\,{\rm mas} \over \pi_{\rm rel}}.
\end{equation}  
The lens would be a sub-Neptune-mass object in the disk or a Saturn-mass object in the bulge. In principle, the lens mass can be constrained by the Bayesian approach based on priors on the Galactic structure (including mass functions) and kinematics. However, we do not conduct a Bayesian analysis because the lens vector proper motion $\muvec_{\rm L} = \muvec_{\rm S} + \muvec_{\rm rel}$, as inferred from the Gaia DR3 vector source proper motion $\muvec_{\rm S}$ and the scalar lens-source relative proper motion $\mu_{\rm rel}$ (dashed circle in Figure~\ref{fig:gaia}), almost equally favors either disk and bulge lenses.

Similar to other short-timescale events, we cannot exclude the presence of a host star at wide separation. We searched possible 2L1S solutions and found that the best-fit 2L1S solution is favored by only $\Delta{\chi^{2}} = 4.7$ over the FSPL solution. For typical disk-lens/bulge-source events with host masses of $M_{\rm host} = 0.6\,M_{\odot}$ and lens distances of $D_{\rm L} = 4.5\,{\rm kpc}$, we have not found any strong evidence for the host within a projected planet-host separation of $5\,{\rm au}$.       

If the lens has a host star, the host light could be visible in the high-resolution images. 
For KMT-2023-BLG-2669, however, we found that based on the results of \citet{mroz24}, the probability for detecting possible lens hosts is very low $(p \sim 17\%$ at $\Delta\theta \sim 200\,{\rm mas})$ due to the bright nature of the source $(K_{\rm S} \backsimeq 13.37)$. 
This implies that it is difficult to distinguish the FFP and wide-orbit planet hypotheses with currently available telescopes.
Nevertheless, because KMT-2023-BLG-2669 is an FSPL FFP, it is possible in principle to search for the host immediately, although such searches are ``high-risk, high-reward''.

\acknowledgments
We are very grateful to the reviewer for useful comments and suggestions, in particular the investigation that is summarized in Section 5. 
We also thank Scott Gaudi for pointing out an error in the original 2L1S-exclusion analysis.
This research has made use of the KMTNet system operated by the Korea Astronomy and Space Science Institute (KASI) at three host sites of CTIO in Chile, SAAO in South Africa, and SSO in Australia. 
Data transfer from the host site to KASI was supported by the Korea Research Environment Open NETwork (KREONET). 
This research was supported by the Korea Astronomy and Space Science Institute under the R\&D program (Project No. 2024-1-832-01) supervised by the Ministry of Science and ICT. 
Work by C.H. was supported by the grants of National Research Foundation of Korea (2019R1A2C2085965 and 2020R1A4A2002885). 
J.C.Y. and I.-G.S. acknowledge support from U.S. NSF Grant No. AST-2108414. 
Y.S. acknowledges support from BSF Grant No. 2020740.
W.Zang. acknowledges the support from the Harvard-Smithsonian Center for Astrophysics through the CfA Fellowship. 
H.Y. and W.Z. acknowledge support by the National Natural Science Foundation of China (Grant No. 12133005).

\begin{deluxetable}{lrr}
\tablecaption{Mean Parameters for FSPL Models}
\tablewidth{0pt}
\tablehead{
\multicolumn{1}{l}{Parameters} &
\multicolumn{1}{c}{FSPL} &
\multicolumn{1}{c}{FSPL $(f_{\rm B} = 0)$}
}
\startdata
$\chi^2/\rm{dof}$             &    1230.93/1223          &     1231.62/1224      \\
$t_0$ $(\rm{HJD}^{\prime})$   & 137.698 $\pm$ 0.005      & 137.698 $\pm$ 0.005   \\
$u_0$                         &   0.508 $\pm$ 0.251      &   0.398 $\pm$ 0.145   \\
$t_{\rm E}$ $(\rm{days})$     &   0.333 $\pm$ 0.033      &   0.334 $\pm$ 0.014   \\
$\rho$                        &   1.456 $\pm$ 0.191      &   1.408 $\pm$ 0.033   \\
$f_{\rm S}$ (KMTS)            &   1.139 $\pm$ 0.269      &   1.040 $\pm$ 0.001   \\
$f_{\rm B}$ (KMTS)            &  -0.099 $\pm$ 0.269      &   --                   
\enddata
\tablecomments{$\rm{HJD}^{\prime} = \rm{HJD} - 2460000$ 
}
\label{tab:FSPL}
\end{deluxetable}

\begin{deluxetable}{lrrr}
\tablecaption{Mean Parameters for 2L1S Models}
\tablewidth{0pt}
\tablehead{
\multicolumn{1}{l}{Parameters} &
\multicolumn{1}{c}{Local A} &
\multicolumn{2}{c}{Local B} \\
\multicolumn{1}{l}{} &
\multicolumn{1}{c}{} &
\multicolumn{1}{c}{free $f_{\rm B}$} &
\multicolumn{1}{c}{$f_{\rm B} = 0$} 
}
\startdata
$\chi^2/\rm{dof}$             &      1230.68/1220          &       1226.95/1220     &       1228.11/1221        \\
$t_0$ $(\rm{HJD}^{\prime})$   &   137.781 $\pm$ 0.306      &   137.698 $\pm$ 0.005  &   137.698 $\pm$ 0.005     \\
$u_0$                         &     0.484 $\pm$ 0.262      &     0.253 $\pm$ 0.174  &     0.135 $\pm$ 0.066     \\
$t_{\rm E}$ $(\rm{days})$     &     0.346 $\pm$ 0.036      &     1.478 $\pm$ 0.574  &     1.215 $\pm$ 0.388     \\
$s$                           &     1.515 $\pm$ 0.372      &    38.593 $\pm$ 23.741 &    33.803 $\pm$ 23.281    \\
$q$ $(10^{-2})$               &   382.209 $\pm$ 248.266    &     0.434 $\pm$ 0.298  &     0.234 $\pm$ 0.104     \\           
$\alpha$                      &     1.966 $\pm$ 1.247      &     3.163 $\pm$ 1.640  &     3.203 $\pm$ 1.639     \\
$\rho$                        &     1.369 $\pm$ 0.192      &     0.439 $\pm$ 0.179  &     0.427 $\pm$ 0.163     \\
$f_{\rm S}$ (KMTS)            &     1.008 $\pm$ 0.260      &     4.196 $\pm$ 3.477  &     1.040 $\pm$ 0.001     \\
$f_{\rm B}$ (KMTS)            &     0.032 $\pm$ 0.260      &    -3.156 $\pm$ 3.477  &     --    
\enddata
\tablecomments{$\rm{HJD}^{\prime} = \rm{HJD} - 2460000$ 
}
\label{tab:2L1S}
\end{deluxetable}

\begin{deluxetable}{lrr}
\tablecaption{Gaia astrometric parameters for the source star}
\tablewidth{0pt}
\tablehead{
\multicolumn{1}{l}{Parameters} &
\multicolumn{1}{c}{DR2} &
\multicolumn{1}{c}{DR3}
}
\startdata
$\varpi~({\rm mas})$                &         --      &    -0.195 $\pm$ 0.352     \\
$\mu_{\alpha}~(\rm{mas\,yr^{-1}})$  &         --      &    -0.458 $\pm$ 0.471     \\
$\mu_{\delta}~(\rm{mas\,yr^{-1}})$  &         --      &    -9.147 $\pm$ 0.253     \\
${\rm RUWE}$                        &         --      &     0.986     \\
$\epsilon$                          &         --      &     0.729     
\enddata
\tablecomments{Parallax $\varpi$, proper motions $(\mu_{\alpha},\,\mu_{\delta})$, 
renormalized unit weight error (RUWE), and astrometric excess noise $\epsilon$ 
come from the second and the third Gaia data release\citep{gaia18,gaia23}. 
We note that the astrometric parameters are not measured in the second data release.
}
\label{tab:gaia}
\end{deluxetable}

\begin{deluxetable}{lllrrrrr}
\tabletypesize{\scriptsize}
\tablecaption{Characteristics of microlensing events from high-resolution imaging}
\tablewidth{0pt}
\tablehead{
\multicolumn{1}{l}{Event} &
\multicolumn{1}{l}{Year} &
\multicolumn{1}{l}{Passband} &
\multicolumn{2}{c}{Magnitude} &
\multicolumn{1}{c}{Contrast Ratio}&
\multicolumn{1}{c}{Separation} &
\multicolumn{1}{c}{Reference} \\
\multicolumn{1}{l}{} &
\multicolumn{1}{l}{} &
\multicolumn{1}{l}{} &
\multicolumn{1}{c}{Source} &
\multicolumn{1}{c}{Lens} &
\multicolumn{1}{c}{} &
\multicolumn{1}{c}{(mas)} &
\multicolumn{1}{c}{} 
}
\startdata
OGLE-2003-BLG-235    &  2018  &  Keck {\it K}      &    17.76 $\pm$ 0.06  &   19.20 $\pm$ 0.11         &  3.77   &  53.50 $\pm$ 0.51  & \citet{bhattacharya23} \\
OGLE-2005-BLG-071    &  2019  &  Keck {\it K}      &    17.68 $\pm$ 0.06  &   18.93 $\pm$ 0.06         &  3.16   &  55.04 $\pm$ 1.30  & \citet{bennett20} \\
OGLE-2005-BLG-169    &  2013  &  Keck {\it H}      &    18.81 $\pm$ 0.08  &   18.20 $\pm$ 0.10         &  1.75   &  61.20 $\pm$ 1.00  & \citet{batista15} \\
		     &  2011  &  HST {\it V}       &    22.21 $\pm$ 0.04  &   22.78 $\pm$ 0.07         &  1.69   &  48.57 $\pm$ 1.29  & \citet{bennett15} \\
		     &  2011  &  HST {\it I}       &    20.56 $\pm$ 0.05  &   20.49 $\pm$ 0.05         &  1.07   &  48.57 $\pm$ 1.29  & \citet{bennett15} \\
MOA-2007-BLG-192     &  2018  &  Keck {\it K}      &    18.94 $\pm$ 0.10  &   18.39 $\pm$ 0.09         &  1.65   &  29.38 $\pm$ 1.46  & \citet{terry24} \\
		     &  2012  &  HST {\it V}       &    24.25 $\pm$ 0.18  &   24.93 $\pm$ 0.32         &  1.87   &  15.75 $\pm$ 6.78  & \citet{terry24} \\
		     &  2012  &  HST {\it I}       &    21.68 $\pm$ 0.16  &   21.56 $\pm$ 0.15         &  1.12   &  18.20 $\pm$ 2.23  & \citet{terry24} \\
		     &  2014  &  HST {\it V}       &    24.25 $\pm$ 0.18  &   24.93 $\pm$ 0.32         &  1.87   &  24.26 $\pm$ 6.22  & \citet{terry24} \\
                     &  2014  &  HST {\it I}       &    21.68 $\pm$ 0.16  &   21.56 $\pm$ 0.15         &  1.12   &  21.84 $\pm$ 1.63  & \citet{terry24} \\
                     &  2023  &  HST {\it I}       &    21.68 $\pm$ 0.16  &   21.56 $\pm$ 0.15         &  1.12   &  43.17 $\pm$ 2.26  & \citet{terry24} \\
MOA-2007-BLG-400     &  2018  &  Keck {\it H}      &    16.58 $\pm$ 0.04  &   19.08 $\pm$ 0.11         & 10.00   &  95.95 $\pm$ 3.46  & \citet{bhattacharya21} \\
		     &  2018  &  Keck {\it K}      &    16.43 $\pm$ 0.04  &   18.93 $\pm$ 0.11         & 10.00   &  96.04 $\pm$ 2.48  & \citet{bhattacharya21} \\
MOA-2008-BLG-379     &  2018  & Keck {\it K}       &    18.87 $\pm$ 0.06  &   18.82 $\pm$ 0.06         &  1.05   &  57.46 $\pm$ 0.61  & \citet{bennett24} \\
		     &  2013  & HST ${V}^{a}$      &    23.67 $\pm$ 0.06  &   $26.49_{-0.66}^{+1.93}$  & 13.43   &  29.71 $\pm$ 0.32  & \citet{bennett24} \\
		     &  2013  & HST {\it I}        &    21.56 $\pm$ 0.15  &   22.75 $\pm$ 0.49         &  2.99   &  29.71 $\pm$ 0.32  & \citet{bennett24} \\
MOA-2009-BLG-319     &  2018  & Keck {\it K}       &    18.12 $\pm$ 0.05  &   19.98 $\pm$ 0.09         &  5.55   &  57.50 $\pm$ 2.40  & \citet{terry21} \\
OGLE-2011-BLG-0950   &  2019  & Keck {\it K}       &    17.02 $\pm$ 0.08  &   16.83 $\pm$ 0.07         &  1.19   &  32.32 $\pm$ 0.79  & \citet{terry22} \\
		     &  2021  & Keck {\it K}       &    17.02 $\pm$ 0.08  &   16.83 $\pm$ 0.07         &  1.19   &  42.18 $\pm$ 0.47  & \citet{terry22} \\
OGLE-2012-BLG-0950   &  2018  & Keck {\it K}       &    17.68 $\pm$ 0.05  &   17.27 $\pm$ 0.04         &  1.46   &  33.34 $\pm$ 2.98  & \citet{bhattacharya18} \\
		     &  2018  & HST ${V}^{b}$      &    20.65 $\pm$ 0.09  &   22.27 $\pm$ 0.21         &  4.45   &  34.53 $\pm$ 2.40  & \citet{bhattacharya18} \\
		     &  2018  & HST {\it I}        &    19.24 $\pm$ 0.06  &   19.57 $\pm$ 0.09         &  1.36   &  34.21 $\pm$ 0.65  & \citet{bhattacharya18} \\
OGLE-2013-BLG-0132   &  2020  & Keck {\it K}       &    17.32 $\pm$ 0.04  &   18.69 $\pm$ 0.04         &  3.53   &  56.91 $\pm$ 0.29  & \citet{rektsini24}  \\
MOA-2013-BLG-220     &  2019  & Keck {\it K}       &    17.20 $\pm$ 0.02  &   17.92 $\pm$ 0.02         &  1.94   &  77.13 $\pm$ 0.45  & \citet{vandorou20}  \\
OGLE-2016-BLG-1195   &  2020  & Keck ${K}^{c}$     &    16.98 $\pm$ 0.05  &   19.96 $\pm$ 0.15         & 15.56   &  54.49 $\pm$ 2.36  & \citet{vandorou23}
\enddata
\tablecomments{(a) For MOA-2008-BLG-379, the HST $V$-band lens magnitude is only marginally detected due to the large uncertainty in the lens flux measurement. See Text. 
(b) The HST $V$-band measurement of OGLE-2012-BLG-0950 provides confirmation of the HST $I$-band and Keck $K$-band measurements. But it cannot be considered as an independent measurement because the lens flux uncertainty is too large. See text. 
(c) For OGLE-2016-BLG-1195, the high-contrast star identified by \citet{vandorou23} is not the lens because the authors identified the wrong source star \citep{gould23}. See Text.
}
\label{tab:high}
\end{deluxetable}

\clearpage

\begin{thebibliography}{99}


\bibitem[Nemiroff \& Wickramasinghe(1994)]{nemiroff94} Nemiroff, R.J., \& Wickramasinghe, W.A.D.T.\ 1994, \apj, 424, L21

\bibitem[Alard \& Lupton(1998)]{alard98} Alard, C. \& Lupton, R. H.\ 1998, \apj, 503, 325

\bibitem[Albrow et al.(2009)]{albrow09} Albrow, M. D., Horne, K., Bramich, D. M., et al.\ 2009, \mnras, 397, 2099

\bibitem[Barclay et al.(2017)]{barclay17} Barclay, T., Quintana, E. V., Raymond, S. N., \& Penny, M. T.\ 2017, \apj, 841, 86

\bibitem[Batista et al.(2015)]{batista15} Batista, V., Beaulieu, J. P., Bennett, D. P., et al.\ 2015, \apj, 808, 170

\bibitem[Bennett \& Rhie(1996)]{bennett96} Bennett, D. P., \& Rhie, S. H.\ 1996, \apj, 472, 660

\bibitem[Bennett et al.(2015)]{bennett15} Bennett, D. P., Bhattacharya, A., Anderson, J., et al.\ 2015, \apj, 808, 169

\bibitem[Bennett et al.(2020)]{bennett20} Bennett, D. P., Bhattacharya, A., Beaulieu, J.-P., et al.\ 2020, \aj, 159, 68

\bibitem[Bennett et al.(2024)]{bennett24} Bennett, D. P., Bhattacharya, A., Beaulieu, J.-P., et al.\ 2024, \aj, 168, 15

\bibitem[Bensby et al.(2013)]{bensby13} Bensby, T. Yee, J.C., Feltzing, S., et al.\ 2013, \aap, 549, A147

\bibitem[Bessell \& Brett(1988)]{bessell88} Bessell, M.S., \& Brett, J. M.\ 1988, \pasp, 100, 1134

\bibitem[Bhattacharya et al.(2018)]{bhattacharya18} Bhattacharya, A., Beaulieu, J. P., Bennett, D. P., et al.\ 2018, \aj, 156, 289

\bibitem[Bhattacharya et al.(2021)]{bhattacharya21} Bhattacharya, A., Bennett, D. P., Beaulieu, J. P., et al.\ 2021, \aj, 162, 60

\bibitem[Bhattacharya et al.(2023)]{bhattacharya23} Bhattacharya, A., Bennett, D. P., Beaulieu, J. P., et al.\ 2023, \aj, 165, 206


\bibitem[Chabrier(2003)]{chabrier03} Chabrier, G.\ 2003, \pasp, 115, 763

\bibitem[Chatterjee et al.(2008)]{chatterjee08} Chatterjee, S., Ford, E. B., Matsumura, S., \& Rasio, F. A.\ 2008, \apj, 686, 580

\bibitem[Gaia Collaboration(2018)]{gaia18} Gaia Collaboration (Brown, A. G. A., et al.)\ 2018, \aap, 616, A1 (Gaia DR2 SI)

\bibitem[Gaia Collaboration(2023)]{gaia23} Gaia Collaboration (Vallenari, A., et al.)\ 2023, \aap, 674, A1 (Gaia DR3 SI)

\bibitem[Gaudi \& Sackett(2000)]{gaudi00} Gaudi, B. S., \& Sackett, P. D.\ 2000, \apj, 528, 56

\bibitem[Gonzalez et al.(2012)]{gonzalez12} Gonzalez, O.~A., Rejkuba, M., Zoccali, M., et al.\ 2012, \aap, 543, A13

\bibitem[Gould(1992)]{gould92} Gould, A. 1992, \apj, 392, 442

\bibitem[Gould(1994)]{gould94} Gould, A. 1994, \apjl, 421, L71

\bibitem[Gould(1997)]{gould97} Gould, A. 1997, \apj, 480, 188  

\bibitem[Gould \& Yee(2013)]{gould13} Gould, A. \& Yee, J. C.\ 2013, \apj, 764, 107

\bibitem[Gould et al.(2022)]{gould22} Gould, A., Jung, Y. K., Hwang, K.-H., et al.\ 2022, JKAS, 55, 173

\bibitem[Gould et al.(2023)]{gould23} Gould, A., Shvartzvald, Y., Zhang, J., et al.\ 2023, \aj, 166, 145

\bibitem[Gould et al.(2024)]{gould24} Gould, A., Yee, J. C. \& Dong, S., 2024, arXiv:2406.14531

\bibitem[Grenman \& Gahm(2014)]{grenman14} Grenman, T. \& Gahm, G. F.\ 2014, \aap, 565, A107

\bibitem[Holz \& Wald(1996)]{holz96} Holz, D.E. \& Wald, R.M. 1996, \apj 471, 64

\bibitem[Kaib et al.(2013)]{kaib13} Kaib, N. A., Raymond, S. N., \& Duncan, M.\ 2013, \nat, 493, 381     

\bibitem[Kim et al.(2016)]{kim16} Kim, S.-L., Lee, C.-U., Park, B.-G., et al.\  2016, JKAS, 49, 37

\bibitem[Kim et al.(2018)]{eventfinder} Kim, D.-J., Kim, H.-W., Hwang, K.-H., et al.\ 2018, \aj, 155, 76

\bibitem[Koshimoto et al.(2023)]{koshimoto23} Koshimoto, N., Sumi, T., Bennett, D. P., et al.\ 2023, \aj, 166, 107

\bibitem[Kervella et al.(2004)]{kervella04} Kervella, P., Th{\'e}venin, F., Di Folco, E., \& S{\'e}gransan, D.\ 2004b, \aap, 426, 297

\bibitem[Kim et al.(2021)]{kim21} Kim, H.-W., Hwang, K.-H., Gould, A., et al.\ 2021, \aj, 162, 15

\bibitem[Ma et al.(2016)]{ma16} Ma, S., Mao, S., Ida, S., Zhu, W., \& Lin, D. N. C.\ 2016, \mnras, 461, L107

\bibitem[Malmberg et al.(2011)]{malmberg11} Malmberg, D., Davies, M. B., \& Heggie, D. C.\ 2011, \mnras, 411, 859

\bibitem[Minniti et al.(2023)]{minniti23} Minniti D., Lucas P., Hempel M., \& The VVV Science Team.\ 2023, yCAT, 2, 376

\bibitem[Mr\'oz et al.(2017)]{mroz17}Mr\'oz, P., Udalski, A., Skowron, J., et al.\ 2017, \nat, 548, 183

\bibitem[Mr\'oz et al.(2018)]{mroz18}Mr\'oz, P., Ryu, Y.-H., Skowron, J., et al.\ 2018, \aj, 155, 121

\bibitem[Mr\'oz et al.(2019a)]{mroz19a}Mr\'oz, P., Udalski, A., Bennett, D. P., et al.\ 2019a, \aap, 622, A201

\bibitem[Mr\'oz et al.(2019b)]{mroz19b}Mr\'oz, P., Udalski, A., Skowron, J., et al. 2019b, \apjs, 244, 29

\bibitem[Mr\'oz et al.(2020a)]{mroz20a}Mr\'oz, P., Poleski, R., Han, C., et al.\ 2020a, \aj, 159, 262

\bibitem[Mr\'oz et al.(2020b)]{mroz20b}Mr\'oz, P., Poleski, R., Gould, A., et al.\ 2020b, \apjl, 903, L11

\bibitem[Mr\'oz et al.(2024)]{mroz24}Mr\'oz, P., Ban, P., Marty, P., \& Poleski, R.\ 2024, \aj, 167, 40

\bibitem[Nataf et al.(2013)]{nataf13} Nataf, D.M., Gould, A., Fouqu\'e, P., et al.\ 2013, \apj, 769, 88

\bibitem[Nemiroff \& Wickramasinghe(1994)]{nemiroff94} Nemiroff, R.J., \& Wickramasinghe, W.A.D.T.\ 1994, \apj, 424, L21

\bibitem[Paczy\'{n}ski(1986)]{paczynski86} Paczy\'{n}ski, B.\ 1986, \apj, 304, 1

\bibitem[Rasio \& Ford(1996)]{rasio96} Rasio, F. A. \& Ford, E. B.\ 1996, Science, 274, 954

\bibitem[Refsdal(1964)]{refsdal64} Refsdal, S. 1964, \mnras, 128, 295

\bibitem[Rektsini et al.(2024)]{rektsini24} Rektsini, N. E., Batista, V., Ranc, C., et al. 2024, \aj, 167, 145

\bibitem[Ryu et al.(2021)]{ryu21} Ryu, Y.-H., Mr\'oz, P., Gould, A., et al.\ 2021, \aj, 161, 126

\bibitem[Smith et al.(2007)]{smith07} Smith, M. C., Woźniak, P., Mao, S., et al.\ 2007, \mnras, 380, 805

\bibitem[Spurzem et al.(2009)]{spurzem09} Spurzem, R., Giersz, M., Heggie, D. C., \& Lin, D. N. C.\ 2009, \apj, 697, 458

\bibitem[Sumi et al.(2011)]{sumi11} Sumi, T., Kamiya, K., Bennett, D. P., et al.\ 2011, \nat, 473, 349

\bibitem[Sumi et al.(2023)]{sumi23} Sumi, T., Koshimoto, N., Bennett, D. P., et al.\ 2023, \aj, 166, 108

\bibitem[Terry et al.(2024)]{terry24} Terry, S. K., Beaulieu, J.-P., Bennett, D. P., et al.\ 2024, arXiv:2403.12118

\bibitem[Terry et al.(2022)]{terry22} Terry, S. K., Bennett, D. P., Bhattacharya, A., et al.\ 2022, \aj, 164, 217
	
\bibitem[Terry et al.(2021)]{terry21} Terry, S. K., Bhattacharya, A., Bennett, D. P., et al.\ 2021, \aj, 161, 54

\bibitem[Tomaney \& Crotts(1996)]{tomaney96} Tomaney, A. B. \& Crotts, A. P. S.\ 1996, \aj, 112, 2872

\bibitem[Udalski et al.(2015)]{udalski15} Udalski, A., Szyma\'{n}ski, M. K., \& Szyma\'{n}ski, G. 2015, Acta Astron, 65, 1

\bibitem[Vandorou et al.(2020)]{vandorou20} Vandorou, A., Bennett, D. P., Beaulieu, J.-P., et al.\ 2020, \aj, 160, 121

\bibitem[Vandorou et al.(2023)]{vandorou23} Vandorou, A., Dang, L., Bennett, D. P., et al.\ 2023, arXiv:2302.01168

\bibitem[Veras et al.(2011)]{veras2011} Veras, D., Wyatt, M. C., Mustill, A. J., Bonsor, A., \& Eldridge, J. J.\ 2011, \mnras, 417, 2104    

\bibitem[Weidenschilling \& Marzari(1996)]{weid96} Weidenschilling, S. J. \& Marzari, F.\ 1996, \nat, 384, 619

\bibitem[Witt \& Mao(1994)]{witt94} Witt, H.J., \& Mao, S.\ 1994, \apj, 429, 66

\bibitem[Whitworth \& Stamatellos(2006)]{whitworth06} Whitworth, A. P. \& Stamatellos, D.\ 2006, \aap, 458, 817

\bibitem[Yang et al.(2024)]{yang24} Yang, H., Yee, J. C., Hwang, K.-H., et al. 2024, \mnras, 528, 11

\bibitem[Yoo et al.(2004)]{yoo04} Yoo, J., DePoy, D. L., Gal-Yam, A., et al.\ 2004, \apj, 603, 139

\end{thebibliography}
\end{document}